\def\Msun{M$_\odot$}

\def\kms    {\ifmmode{{\rm \ts km\ts s}^{-1}}\else{\ts km\ts s$^{-1}$}\fi}
\def\LprimeCO{{\hbox {$L^\prime_{\rm CO}$}}}

\def \Kkmspc{K\,\kms\,pc$^2$}
\documentclass{aa}  
\usepackage{graphicx}
\usepackage{txfonts}
\begin{document}
   \title{Observations of CO in the eastern filaments of NGC 1275}

   \author{P. Salom\'e
          \inst{1}
          \and
          Y. Revaz\inst{2,5}
          \and
          F. Combes\inst{2}
          \and
          J. Pety\inst{1,2}
          \and
          D. Downes\inst{1}
          \and
          A.C. Edge\inst{3}
          \and
          A.C. Fabian\inst{4}
}


   \institute{Institut de Radio Astronomie Millim\'etrique, Domaine
     Universitaire, 38406 St.\ Martin d'H\`eres, France
         \and
             LERMA, Observatoire de Paris, 61 av.\ de l'Observatoire, 
             75014 Paris, France
         \and
             Department of Physics, University of Durham, South Road,
             Durham, DH1 3LEi, UK
         \and
             IoA, Madingley Road, Cambridge, CB3 OHA, UK
         \and             
               Laboratoire d'Astrophysique, \'Ecole Polytechnique 
             F\'ed\'erale de Lausanne (EPFL), Observatoire, 1290 
             Sauverny, Switzerland}

   \date{Received January 17th, 2008; accepted March 21st, 2008}

\abstract{We recently found extended CO(2--1)
emission from cold molecular gas embedded in the network of H$\alpha$
filaments surrounding the galaxy NGC~1275 (Salome et al.\ 2006).  We
now present CO(2--1) interferometer maps of the eastern filaments, at
high spatial and spectral resolutions. 

The cold molecular gas is
detected by the Plateau de Bure Interferometer along the eastern
filaments over an extent of 15$''$, or with a projected length of
5\,kpc. In our 2.5$''$ beam, the main CO filament is mostly unresolved
along its minor axis. The multiple peaks along the CO filaments and
the low values of the observed CO brightness temperatures imply
further unresolved structures that may be giant molecular clouds. These
clouds have very narrow line-width emission lines ($\sim$30 km/s).
The CO emission is optically thick. It very likely traces cold clouds
bound under their own self-gravity that may be falling back in the
gravitational potential well of the galaxy. Such a picture would
agree with current models of ``positive feedback'' in which some 
of the hot gas around NGC 1275 (a) is trapped by buoyantly rising
bubbles inflated by the energy input of the 3C84 AGN, (b) subsequently
cools efficiently at a larger radius around the edges of the hot
bubbles, and (c) then falls back in self-gravitating clouds of
molecular gas toward the center of the galaxy.

\keywords{galaxies: cooling flows  --- galaxies: individual: NGC 1275
      --- galaxies: ISM, intergalactic medium --- galaxies: kinematics and dynamics}
}
   \maketitle
%
\section{Introduction}
The galaxy NGC~1275 lies at the center of the Perseus cluster, the
brightest galaxy cluster in the sky in X-rays. 
Its surface brightness follows a $r^{-1/4}$ law with an
effective radius of 15\,kpc. This starlight dominates over the
light of the Perseus cluster right up to the Holmberg radius of
NGC~1275 of 122\,kpc.  Beyond this distance, the cD envelope of NGC~1275
continues out to a cutoff radius of 250\,kpc (Oemler 1976; Schombert
1987; Rebusco et al.\ 2005\footnote {All radii are re-scaled with
$H_0$ = 71\,\kms\,Mpc$^{-1}$.}). 
The galaxy has a remarkable system
of H$\alpha$ filaments with a broad range of gas temperatures and
densities (e.g., Crawford et al.\ 1999; Conselice et al.\ 2001). 
The filaments are very luminous (4 $\times$ 10$^{42}$ erg \, s$^{-1}$ 
in H$\alpha$ and [NII]; Heckman et al., 1989), but the gas 
excitation mechanism is still a matter of debate. The filaments do not contain 
many young star clusters to ionize the gas; a constant source of 
heating could be present that produces their low-excitation spectra, 
but AGN-heating of the filaments has been ruled out. 
Hot (2000\,K) molecular gas is also detected in the filament system in
the H$_2$ ro-vibrational lines at 2\,$\mu$m (e.g., Hatch et al. 2005), and
cooler (200 K) molecular gas is detected in the mid-IR H$_2$ rotational
lines (Johnstone et al. 2007).  Evidence for possibly even colder (10
to 100\,K) molecular gas has recently been found in the filaments in
CO (Salome et al. 2006; 2008, Lim et al. 2008). 
The origin of the filaments is unknown; they may be related to the
cooling of uplifted gas.  A currently popular scenario is that the
supermassive black hole 3C84 emits radio jets surrounded by broad
envelopes of entrained, hot, fast-moving but non-relativistic plasma,
that pushes and compresses the surrounding gas within the
``isothermal'' core of NGC 1275.  The resulting bubbles
then expand in the surrounding gas and rise higher in the central
galaxy's potential well (for recent model simulations see, e.g.,
Br\"uggen $\&$ Kaiser 2002; Churazov et al.\ 2000; Soker $\&$
Pizzolato 2005; Reynolds et al. 2005; Revaz et al.\ 2008; Mathews \&
Brighenti 2007; Sutherland \& Bicknell 2007, Vernaleo \& Reynolds 2007).  
Some of the lower-temperature, higher-density ambient medium is
carried by the outward moving bubbles and slowly cools at greater
radius. The further decrease of its pressure and the increase of its
density lead to runaway cooling instabilities that generate long
filamentary structures. The cooled gas falls back down in the
potential well of the central galaxy, on a time scale of 10$^8$ years
(Revaz et al.\ 2008). If the cooling gas is able to condense into
giant molecular clouds, they should be able to form new star clusters
on their way back in. In NGC~1275, the H$\alpha$ kinematics support this
scenario: the measured velocity along the filaments change sign in the
middle, and Hatch et al (2006) deduce that the gas is inflowing at the
base, while outflowing at the top, with an interruption in the middle
coinciding with an X-ray shock and a ghost bubble. 

\begin{figure*}
\begin{center}
\includegraphics[width=11cm, angle=-90]{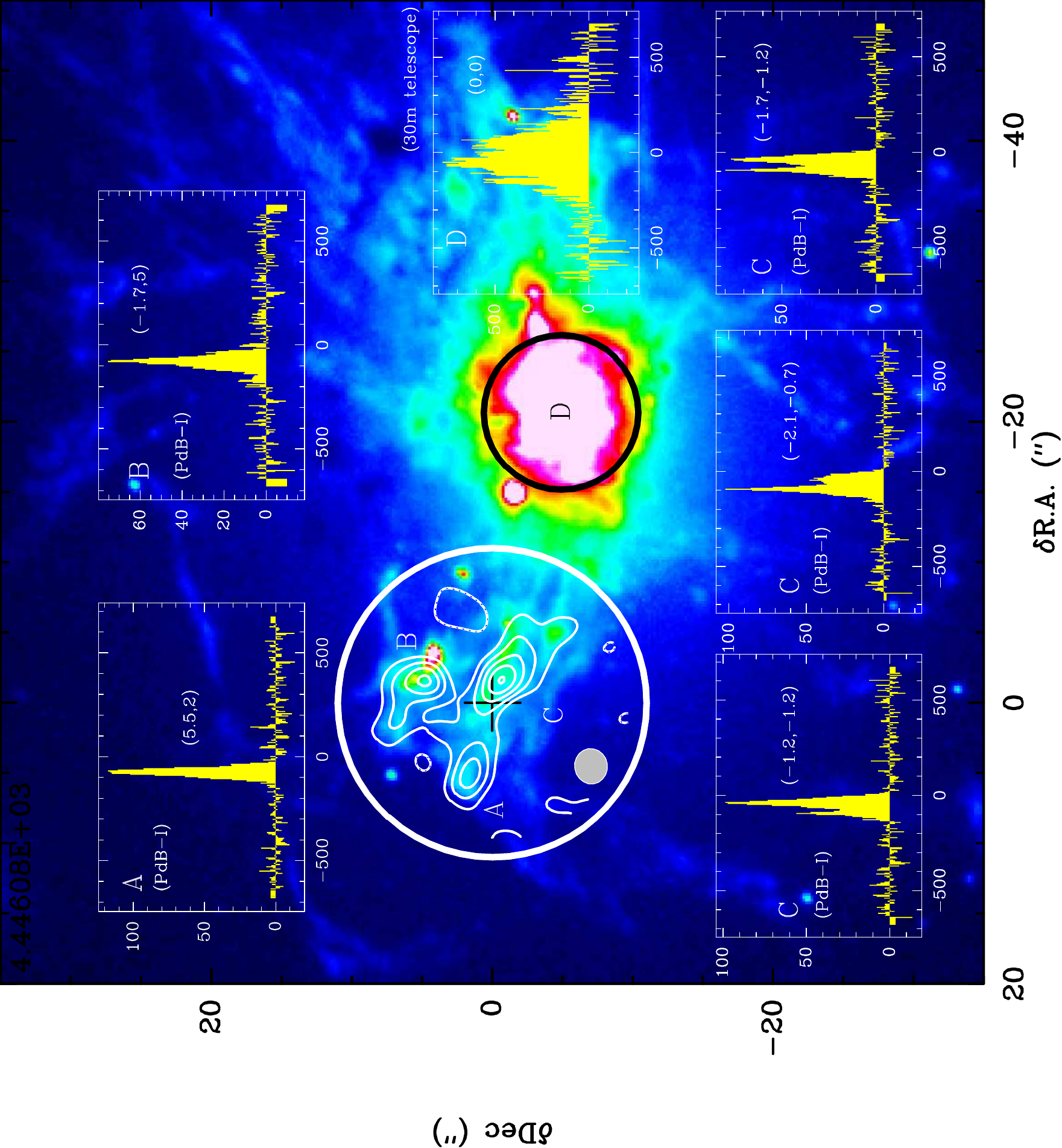}
\includegraphics[width=11cm, angle=-90]{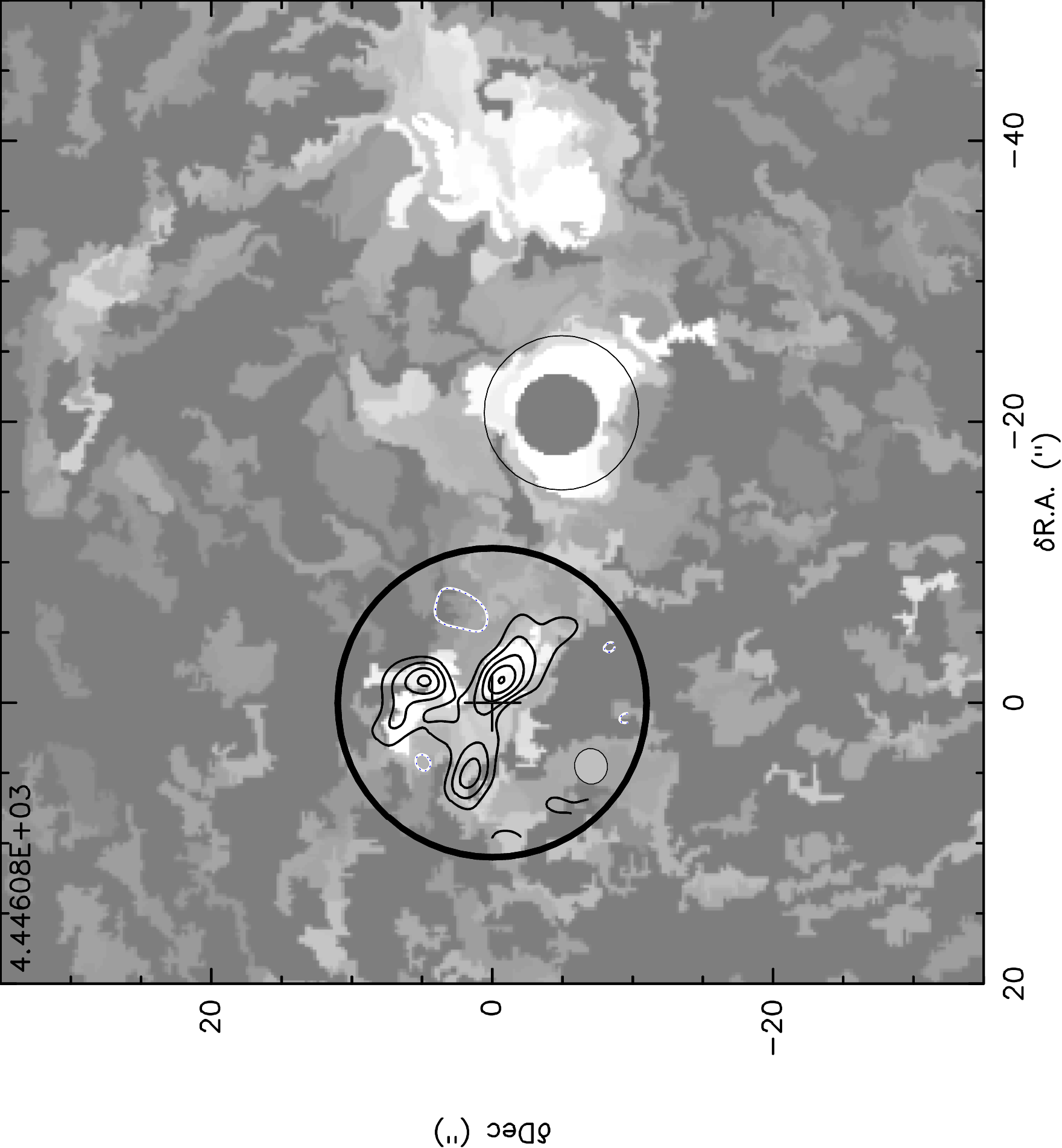}
\caption{{\bf Top:} Integrated CO(2--1) intensity (contours) superposed on 
an H$\alpha$ image (Conselice et al., 2001) of NGC 1275.  The
inteferometer data were averaged over velocity offsets from $-$180 to
$-$5\,\kms .  Contour steps are 1.6\,Jy\,\kms\,beam$^{-1}$.  The white
circle shows the interferometer primary beam, and the grey ellipse at
lower left, within the circle, shows the synthesized beam of
$2.7''\times 2.5''$ for the merged single-dish+inteferometer data.
The CO spectra are from single pixels in regions A, B and C. The
CO(2--1) spectrum of region D, the center of NGC~1275, is from the
single-dish data only, and indicates an even larger amount of
molecular gas in the galaxy's center, with a much greater line width
than in the filaments.  Intensities of the spectra are in mJy in the
2.7$''$ beam, except for region D, which is in units of mJy in the
single-dish 11$''$ beam (black circle around region D).  Velocity
offsets are relative to 226.559\,GHz. {\bf Bottom:} Integrated CO(2--1)
intensity superposed on the hot gas mass distribution derived from 0.5
keV X-ray observations of NGC 1275, from Fabian et al. (2006). CO
contours, primary beam and synthesized beam are as marked in the upper
diagram.}
\label{merged-sum}
\end{center}
\end{figure*}

This paper presents CO(2--1) images of the eastern filaments located
at a projected distance of 8\,kpc from the center of NGC~1275. Sections
2 and 3 report the high spatial and spectral resolution observations
of the molecular gas morphology and kinematics. Section 4 compares these
results with bubble models, and section 5 discusses possible
explanations for the origin of the cold gas in the filaments.

\begin{figure} \centering
\begin{tabular}{l}
\includegraphics[width=5.1cm, angle=-90]{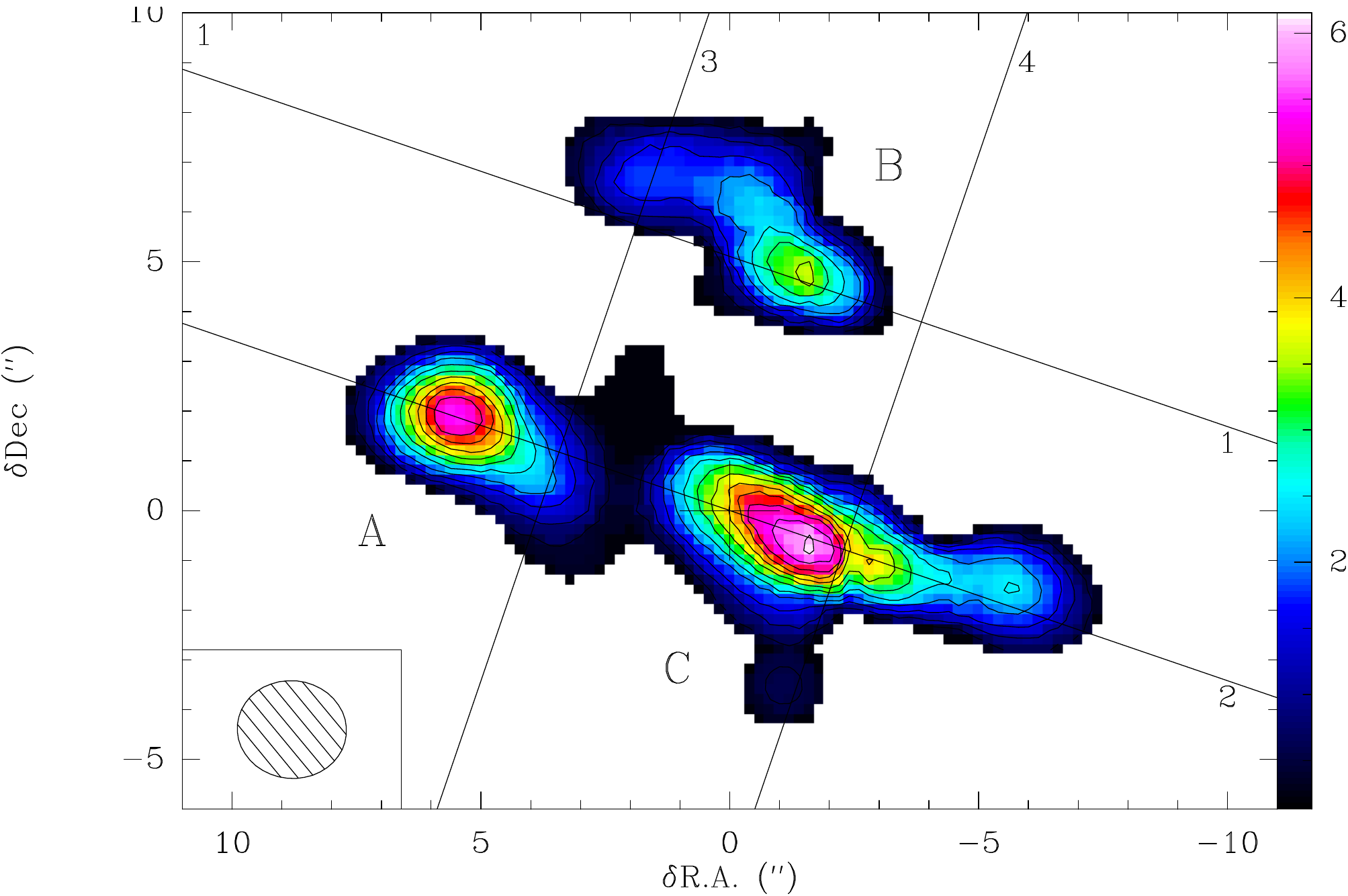}\\ 
\\
\includegraphics[width=5.1cm, angle=-90]{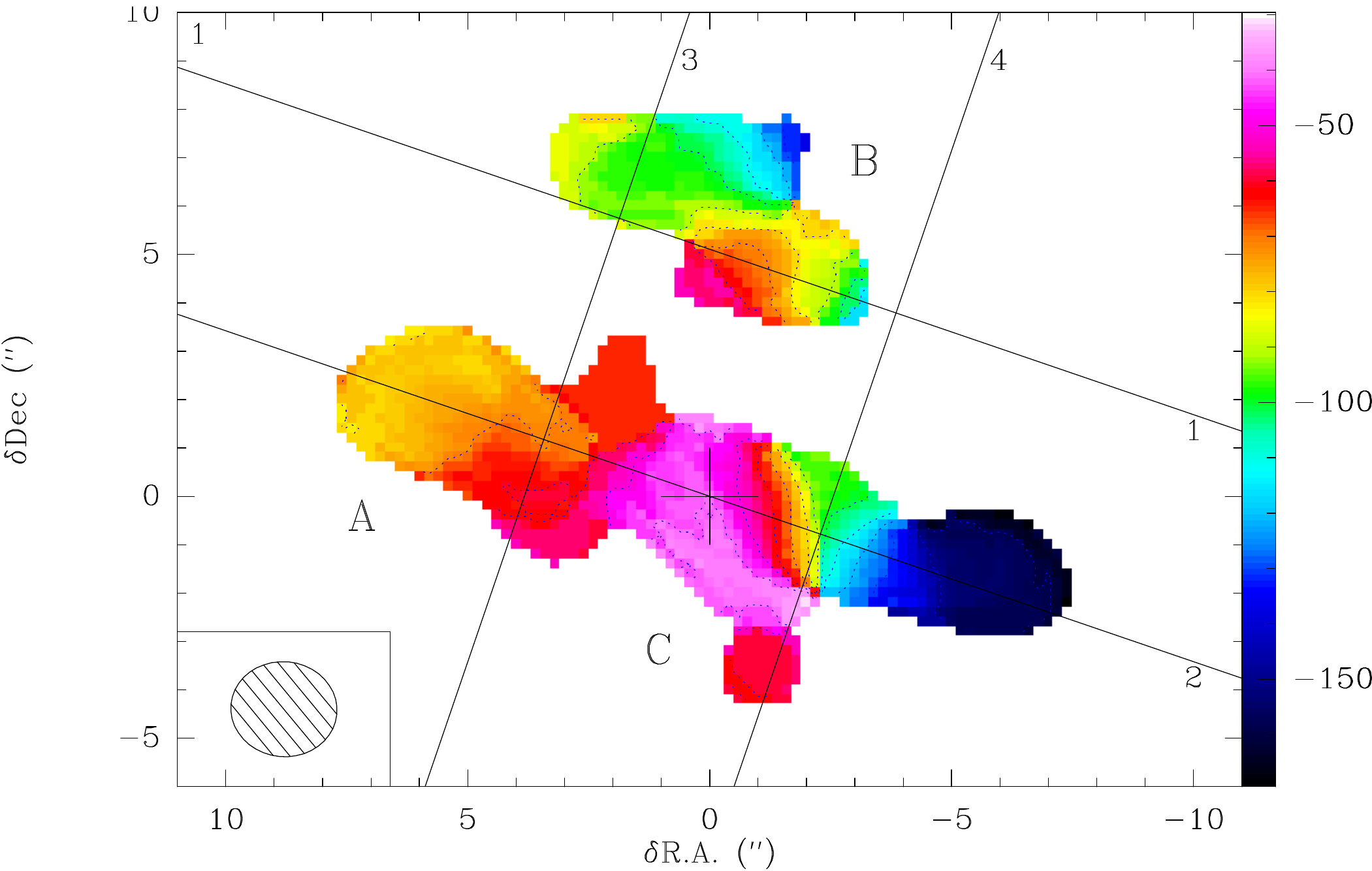}\\
\end{tabular}
\caption{{\it Upper:} Mean CO(2-1) intensity along the filament, with 
contour steps of 0.5\,Jy\,\kms\,beam$^{-1}$.  {\it Lower:} Isovelocity
contours, in steps of 10\,\kms .  The lines labeled 1 to 4 mark the
position-velocity slices in Fig \ref{xv2}. The beam (lower left) is
$2.2''\times 2.0''$).}
\label{xv1} 
\end{figure} 
%
\begin{figure} \centering
\begin{tabular}{r}
\includegraphics[width=9.3cm, angle=-90]{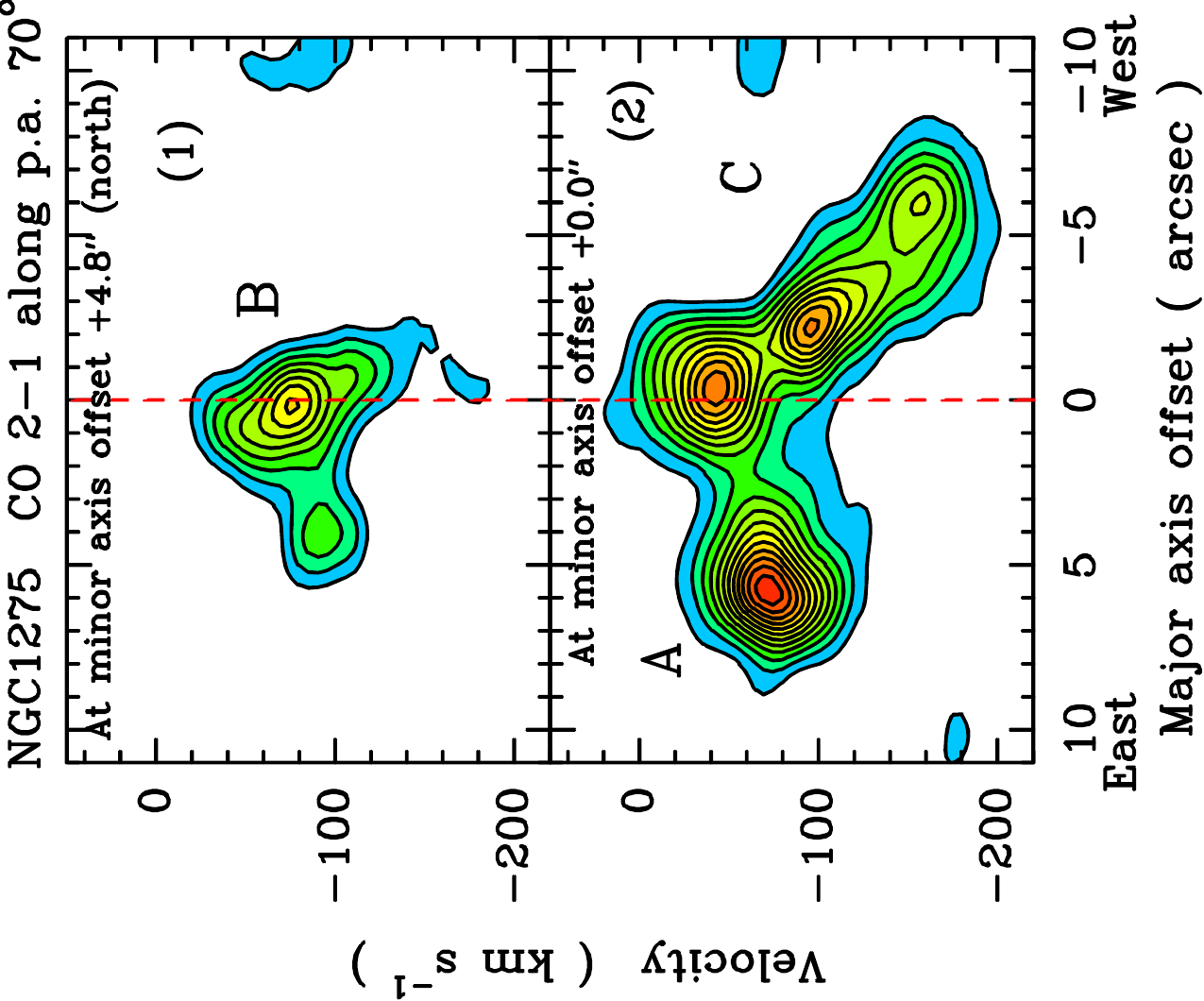} \\
\\
\includegraphics[width=9.3cm, angle=-90]{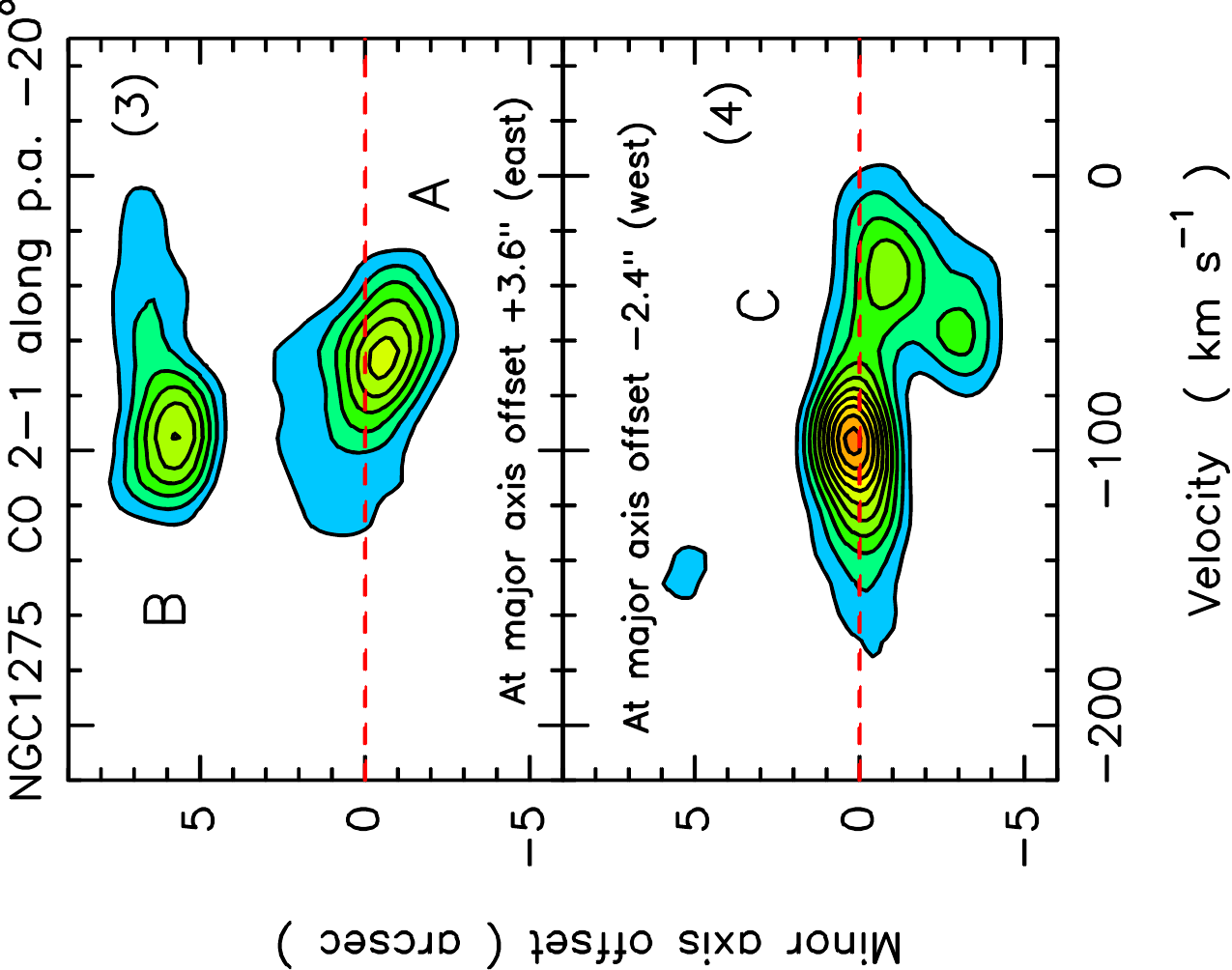} \\
\end{tabular}
\caption{P-V plots of CO(2-1) parallel to the
CO filament's major axis (p.a.\ +70$^\circ$; upper panels) and minor
axis (p.a.\ $-20^\circ$; lower panels).  Panel numbers 1 to 4
correspond to the lines in Fig.~1. Labels A,B,C indicate the CO peaks
(Figs~1 and 3). Contours are by 8 mJy/beam. }
\label{xv2} 
\end{figure}

\section{Observations}
To investigate in more detail the structure of the cool molecular gas
associated with the H$\alpha$ filament system of NGC~1275, we made new
CO(2--1) observations of the prominent eastern filament located at
($\Delta\alpha ,\Delta\delta$) = ($+20'', +10''$) from the radio
source 3C84 at the galaxy's center.  We used the IRAM Plateau de Bure
Interferometer with 6 antennas in the most compact configuration,
which gave beams of 2.5$''\times 2.4''$ (natural weighting) and
$2.2''\times 2.0''$ (robust weighting).  The receivers observed the H
and V polarizations simultaneously at 226.559\,GHz, which is equal to
the rest frequency of the CO(2-1) line divided by (1+$z_{\rm lsr}$),
where we took $z_{\rm lsr}$ = 0.01756, as the reference redshift for
the CO in the center of NGC~1275.  All velocity offsets in the figures
are relative to this frequency.  The SSB system temperatures were
$\sim$220\,K at 1.3\,mm. The 1 GHz-wide IF band covered a velocity
range of 1354\,km\,s$^{-1}$, and was observed with a channel spacing
of 2.5\,MHz.  Fluxes were calibrated with MWC349 and reference
quasars. The total on-source integration time with 6 antenna was
6.4$^{\rm h}$.  Amplitudes and phases were calibrated with 3C84 whose
flux density was 4.3\,Jy at 1.3\,mm.  During the observations, the antennas
were pointed toward the eastern filament, and the position offsets in
the figures are relative to a (0,0) position of 03$^{\rm h}$19$^{\rm
m}$50.0$^{\rm s}$, +41$^\circ$ 30$^{'}$ 47.0$^{''}$ (J2000), centered
on this filament.  Because 3C84 was a full primary beamwidth (22$''$)
away from this position, its apparent flux was attenuated to 97\,mJy
on our maps, thereby reducing the effect of beam sidelobe responses to
this strong continuum source.
 
\section{Filament-like structure in the CO} 
\subsection{Overall morphology of the CO filaments}
Figure \ref{xv1} shows the mean CO(2--1) intensity in the east
filament region, and the moment map of isovelocity contours.  The
molecular gas is in an elongated structure that coincides with the
H$\alpha$ filaments (see e.g., the H$\alpha$ images by Conselice et
al.\ 2001).  All the CO(2--1) flux found with the 30m telescope in the
eastern filament region is detected by the interferometer:
40\,Jy\,\kms. This means that {\itshape all the molecular gas in the
interferometer field of view is in this elongated structure --- no
further extended emission exists} that has been lost by missing short
spacings of the interferometer.  Much more cold molecular gas is
present within the central 2-kpc radius, as is shown by the spectrum of CO
emission on the position of 3C84 (Fig.~3, region D; see Salom\'e
et al.\ 2008 for details).

\begin{table*}
\caption{CO clumps in the eastern filament region. 
} 
\centering
\begin{tabular}{cccccccccc}
\hline \hline
&\multicolumn{2}{c}{Position offsets} &Velocity
&Peak line  &\multicolumn{2}{c}{Smoothed size}
&linewidth  &CO luminosity  &Gas \\
Clump &R.A.     &Dec.     &offset      
&flux      &$\Delta$R.A.   &$\Delta$Dec.  
&FWHM    &\LprimeCO   &Mass \\  
number    
&(arcsec)  &(arcsec)  &(\kms ) 
&(Jy beam$^{-1}$)  &(arcsec)    &(arcsec)     
&(\kms )  &(10$^7$\,\Kkmspc ) &(10$^7$\,\Msun )\\
\hline            
1   &$+5.2$  &$+1.5$  &$-63.6$   &  0.16  &  2.5  &  3.3 & 29 &2.1 &9.6\\
2   &$+1.2$  &$+6.3$  &$-84.9$   &  0.13  &  4.6  &  2.5 & 15 &1.2 &5.6\\
3   &$-1.9$  &$+4.4$  &$-79.3$   &  0.12  &  2.5  &  2.5 & 17 &0.7 &3.3\\
4   &$-0.5$  &$-0.4$  &$-37.1$   &  0.11  &  2.5  &  4.0 & 25 &1.5 &6.9 \\
5   &$-2.0$  &$+0.0$  &$-89.9$   &  0.11  &  2.5  &  3.1 & 31 &1.4 &6.5\\
6   &$+5.8$  &$+2.1$  &$-84.7$   &  0.10  &  2.5  &  2.6 & 23 &0.8 &3.6\\
7   &$-5.3$  &$-1.7$  &$-148.2$  &  0.10  &  2.7  &  3.7 & 31 &1.6 &7.6\\
\hline     
\multicolumn{10}{l}{Estimated errors are $\pm 20$\% in all quantities.
Clump sizes (FWHM) are convolved with the 2.5$''$ beam.}
\\  
\multicolumn{10}{l}{Gas masses (H$_2$+He) are 
for solar abundances (as for iron-line X-rays) and $M/L^\prime_{\rm CO}$ = 
4.6\,M$_\odot$\,(\Kkmspc )$^{-1}$.}       
\\
\multicolumn{10}{l}{Clumps 1 and 6 are within region A but have different 
velocities, and clumps 4 and 5 are within region C (Fig.1).}\\
\end{tabular}
\label{parameters} 
\end{table*}

\begin{figure*}
\centering
\includegraphics[width=14cm, angle=-90]{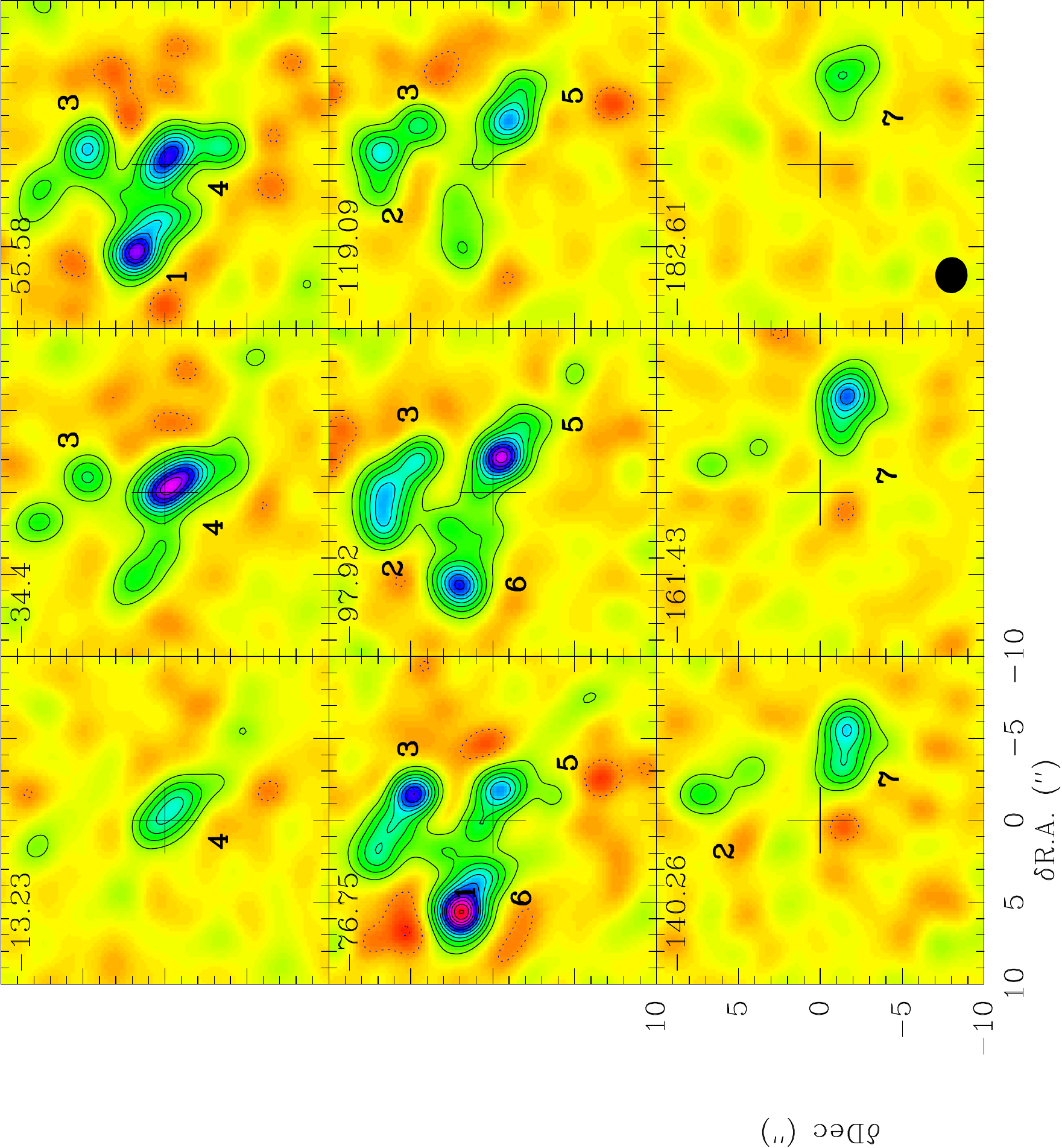}
\caption{
CO(2--1) maps, centered on the eastern filaments of NGC 1275, and
binned in 21.2\,\kms\ channels to show the main clumps, numbered as in
Table~1.  Contour steps are 8.9\,mJy\,beam$^{-1}$ (4$\sigma$).  The
beam is $2.2''\times 2.0''$.  Position offsets are relative to the map
center at 03$^{\rm h}$19$^{\rm m}$50.0$^{\rm s}$, +41$^\circ$ 30$^{'}$
47.0$^{''}$ (J2000), marked with a cross.  Velocity offsets (upper
left corners) are relative to 226.559\,GHz.  }
\label{over-chan} 
\end{figure*} 

Figure~\ref{xv2} presents position-velocity diagrams along the
main filament's major axis (p.a.\ 70$^\circ$), and along its minor
axis (p.a.\ $-20^\circ$).  Region B is at $-$70\kms with a hint of a
velocity gradient. Region A also peaks at $-$70\kms . Region C is
resolved in 3 substructures with a clear velocity change from
$-$160\,\kms\ at offset $-6''$ to $-$100\,\kms\ at offset $-$2$''$ and
then to $-$40\,\kms\ at offset 0$''$.  The velocity change of
120\,\kms\ over a projected extent of 6$''$, or 2.1\,kpc, gives an
apparent gradient in the plane of the sky of the line-of-sight
velocity of 0.06\,\kms\,pc$^{-1}$.
 The p-v diagrams parallel to the minor axis Fig.~2, panels 3
\& 4) show no clear velocity gradient. Regions A, B and C may
simply be three separate structures.  The CO filaments are seen
projected on the plane of the sky, and they are likely to be made of
different sub-filaments, as in the H$\alpha$ images.

The interferometer maps show several substructures inside the
filaments. We merged the new interferometer data with our previous
CO(2--1) observations at the IRAM 30\,m telescope (Salom\'e et al.,
2006) that covered a much larger field of view, in order to improve
the cleaning algorithm.  Figure~\ref{merged-sum}, shows the merged
single-dish+interferometer map, with insets of five spectra at the CO
peaks in the filaments.  The CO(2--1) lines in the filaments are
narrow ($\le$100 \kms ). The mean velocity of the CO is shifted by
about $-$100 \kms\ relative to the midpoint of the CO(2--1) line at
the center of NGC 1275, which is at a velocity offset of $-$30\,\kms ,
and is also much broader (FWHP $\sim$300 \kms ) than the CO lines in
the filaments.  The peak of the central CO emission is 7 to 10 times
the CO peak intensity in the brightest components of the eastern
filament system.

All the filament regions labeled A,B, and C are elongated.  The long
axes are resolved by the interferometer, while the apparent minor axes
are close to the beam size, indicating they are unresolved, with sizes
$<1.5''$ ($<500$\,pc).  The apparent (beam-smoothed) sizes for region
A are 7$''$$\times$3$''$, for region B, 7$''$$\times$$3''$, and for
region C, 10$''$$\times$3$''$.  Regions A and C are aligned on the
same H$\alpha$ filament, while region B lies on a separate H$\alpha$
filament, $5''$ north of A-C.  These filaments are projected on the
plane of the sky, so the minimum length of the region C major axis in
CO is 3.5\,kpc.  If regions A and C are two substructures of 
the same CO filament, then its projected length could be as much as 
6\,kpc.

\subsection{Are the CO clumps complexes of GMCs?}
Figure~\ref{over-chan} shows the CO(2--1) maps binned in
21\,\kms\ channels to emphasize the main clumps.  We used the program
GAUSSCLUMPS (Stutzki \& G\"usten 1990; Kramer et al.\ 1998) to fit
gaussians to the substructures inside the CO filaments.  Table
\ref{parameters} summarizes the results of the fits.  The clumps have
apparent sizes (convolved with the 2.5$''$ beam) of 2.5$''$
to 4.6$''$.  While some of the clumps are slightly resolved along
their major axis, all of them are unresolved along their minor axis,
and our upper limit on their deconvolved width is about half
the beamwidth, or 1.3$''$ (450\,pc).  To estimate the mass, we derived
the CO line luminosity, $L^{\prime}_{\rm CO}$, with the formula of Solomon
et al.\ (1997), and then applied a standard conversion factor of 4.6
M$_\odot$\,(\Kkmspc )$^{-1}$.  With this factor, the estimated gas
mass of each clump is $\sim$5$\times$10$^7$\,\Msun , and the total gas mass
(H$_2$ + He) in the CO east filament region is $4\times 10^8$\,\Msun .

The observed brightness temperatures of the clumps are 0.39 to
0.63\,K, which is one to two orders of magnitude too low for normal,
optically thick molecular clouds. Evidence that this CO is indeed
optically thick comes from the observations with the IRAM 30m
telescope by Salom\'e et al. (2008).  A comparison of the 30m CO(1-0)
and (2-1) brightness temperatures convolved to the same beamsize
(22$''$) and centered on the eastern filament region gave a line ratio
close to unity, consistent with optically thick line radiation.  One
possibility to reconcile the low observed brightness temperatures with
their much higher expected values is that each clump is actually a
complex of several Giant Molecular Clouds (GMCs).  For typical Milky
Way GMCs, the CO brightness temperatures are 20 to 40\,K, the gas mass
is 10$^6$\,\Msun , and the typical radius is 30\,pc (0.1$''$ at the
distance of NGC~1275).  If there were 10 such GMCs in each of the
clumps on our maps, then the area filling factor within our 2.5$''$
beam would be f$_a\sim$1 to 2\% .  For GMCs with such a mass and
radius, the volume densities are several hundred H$_2$ molecules
cm$^{-3}$, and the surface densities are $\sim$10$^{22}$\,cm$^{-2}$.
For comparison, the gas masses and extents of Milky Way-type GMCs are
similar to the mass and size of the star cluster found in the Eastern
filament by Shields \& Filipenko (1990), so the CO clumps on our maps
could conceivably be the birth sites of star clusters.  It remains to
be explain why only one young optical star cluster, with HII region,
is seen among all of these presumed GMC associations, and so why the
rate of star formation is unusually low. The molecular gas may not be
in GMCs at all. It would then not be dense enough to form
stars. Higher spatial resolution observations are planned to look for
smaller structures inside the eastern filament. The presence of
magnetic field inside the filament might also help to prevent the
molecular gas to be gravitationally unstable and thereby decrease the
star formation efficiency. Magnetic pressure would also help to
confine the gas inside those puzzling extremely long and thin
filaments.
 
\section{Comparison with bubble models}

These CO observations show a complicated velocity pattern in the
filament that extends from region A through region C (Fig.1, upper).
From region A to the zero position, the transverse gradient in the
line-of-sight velocity is positive, with a value of +40 km/s / 2.1
kpc. Over region C, the velocity gradient is negative with a value of
$-$120 km/s / 2.1 kpc. If these sections are continuous structures, the
apparent transverse gradients are +0.02 km/s/pc and $-$0.06 km/s/pc,
comparable to velocity gradients through molecular clouds or along
spiral arms in our Galaxy. Similar large-scale motions of
$\sim100$\,\kms over distances $\sim 20$ kpc have been observed in
H$_2$ and P$\alpha$ lines by Jaffe et al. (2005) from BCGs in other
clusters of galaxies.  If the eastern filament is on the front side of
NGC~1275, then the velocity shifts, which are blue-shifted relative to
the systemic velocity, indicate outflowing material, as deduced for
the H$\alpha$ filaments by Hatch et al. (2006).  If the eastern
filament is on the rear side of the galaxy, then the blue-shifted CO
indicates infalling material along the filament.

In the latter case, the inflow may be explained by the model proposed
by Revaz et al. (2008), in which the ``heavy'' giant molecular clouds
(GMCs) fall back in a thin stream on the central axis of the
buoyantly-rising bubble.  According to this model high temperature
filaments of ionized gas resulting from uplifted ambient hot gas could
have cooled and formed the observed eastern filament.
Once the plasma temperature drops below $10^6$\,K, the gas is no
longer pressure-supported and falls back towards the center.  This
scenario naturally reproduces part of the observed velocity gradient,
because cold gas emerges from higher-temperature gas at the top of the
filament with nearly zero velocity offset relative to the systemic
velocity of NGC~1275, while cold gas at the bottom of the filament has
sped up by falling in.  Figure 1 (position-position diagram) as well
as Figure 2 (position-velocity) may be qualitatively reproduced (see
Figure 5) by the model 2 of Revaz et al. (2008) at $t=250\,\rm{Myr}$.
Figure 5 assumes that the cold filament is behind the galaxy (to
produce a net blue shift), and being viewed down its length, i.e.,
observed at an angle of 10$^\circ$ from our line of sight.  In this
model, the cold gas is produced between 25 and 50\,kpc, corresponding
to a projected observed distance of 5\,kpc to 10\,kpc. A velocity
gradient larger than $100\,\rm{\kms }$ is easily reproduced.  The mass
enclosed in the red box is equal $3\times 10^8$\,\Msun, similar to the
$4\times 10^8$\,\Msun\ observed in CO.  If the viewing angle along the
line of sight is larger than $45^\circ$, the velocity gradient is
reduced and the data are harder to reproduce.  In this scenario, the
velocity gradient also tells us that the cold gas, if free falling,
may not have been created too far away from the center, otherwise, the
maximal velocity (at the bottom of the filament) should be much higher
than observed. It is possible, though, that GMC are globally subject
to ram pressure on the hot gas, and are somewhat braked in their
free-fall, dragging the dense clumps gravitationally locked in. Magnetic 
fields might also play an important role in holding
clouds together.

\begin{figure}
\resizebox{\hsize}{!}{\includegraphics[angle=-90]{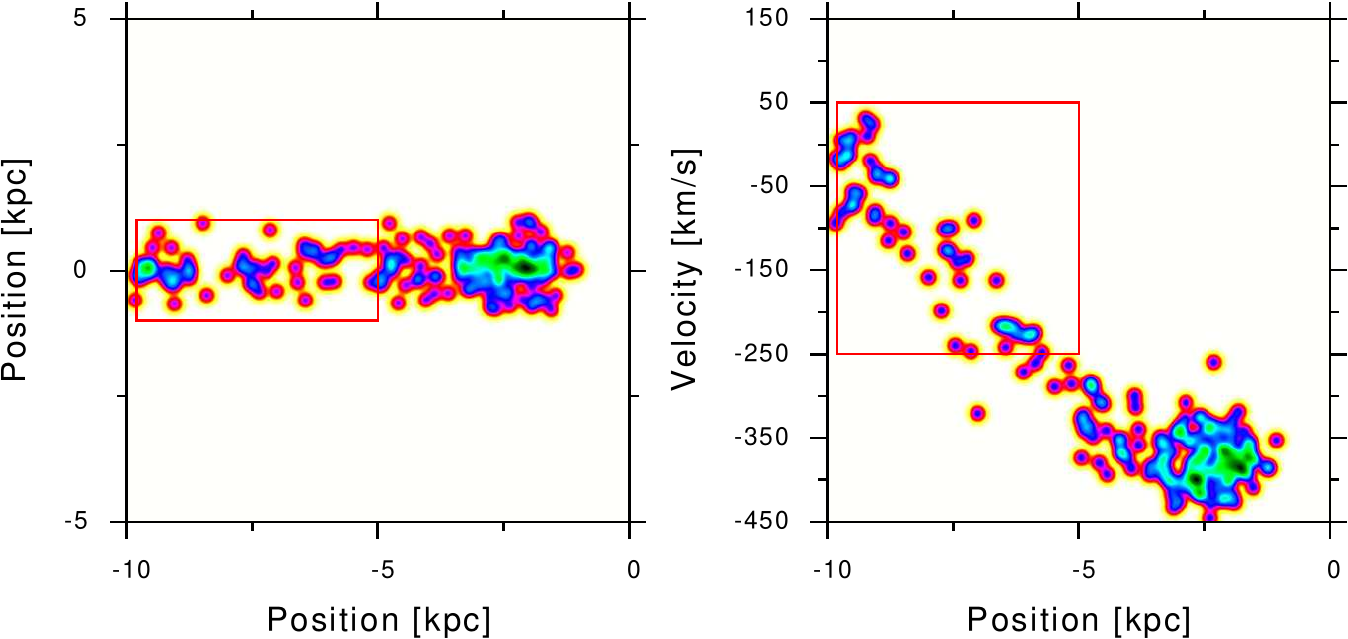}}
\caption{{\it Left:} Surface density of cool gas  
in simulated filaments in model 2 of Revaz et al.\ (2008), at
$t=250\,\rm{Myr}$. The model filament is inclined at 10$^\circ$ to the
observer's line of sight.  {\it Right:} position-velocity diagram for
the same filament.  The red box shows a p-v sector in the model that
is comparable to that of the eastern filament region observed in CO.}
\label{fig0}
\end{figure}
\section{Origin of the Molecular Gas}
The cold molecular filament mapped here is part of an extended ($\sim$ 30 kpc)
network of filamentary structures seen in H$\alpha$ and soft X-ray
emission. No such filamentary structure is ever seen in galaxy
interactions, but is specific to {\itshape{cooling flow galaxies}}. There is
a good correspondence between ionized gas and molecular gas
in the filaments, and also a very good correspondence
between the morphologies of the CO emission and the gas that radiates at higher
temperature in X-rays (Fig. 3). 
The single-dish map from the IRAM 30m telescope (Salom\'e et al.
2006, their Fig.~5) showed no sign of rotation of the molecular gas
around the center of NGC~1275, on scales of $\sim 60''$ (20 kpc).  The CO lines
have negative velocities on both sides of the galaxy's center.  
This surprising pattern has also been confirmed by CO(2-1) mapping 
with the SMA (Lim et al, 2008). This
molecular gas was interpreted as being either flowing toward the
central object or outflowing in a wake of an expanding bubble. The
lack of a rotation pattern on these scales makes it unlikely to be gas tidally
stripped from another galaxy. It also rules out a large amount of 
rotationally-supported cold gas in the galaxy potential well.
In our interpretation the gas dragged outwards by the rising bubble is 
first hot, and cools progressively.  
We think that the molecular gas detected in CO is associated with the 
cooled gas that may be partly falling back on the galaxy center.

The lack of any rotation pattern closer to the center is puzzling and
also supports a scheme where the filament network is spread all around
NGC~1275. If we assume that the cold gas is in free fall in the galaxy
potential well, then a model of cold gas flowing down towards the galaxy
from all directions could explain the absence of global angular
momentum and prevent any rotation closer to the center.  The filaments
are not all radial, but are perturbed and asymmetric, with a non-zero
impact parameter. In isolation, any one filament would then give rise to a
certain rotation.  But the infall of several filaments in any
direction cancels out the net angular momentum. The large velocity
dispersion in the central region (Region D in Fig \ref{merged-sum})
supports this scenario. When compared to the very narrow spectra of
region A, B and C, a natural explanation is that the center could be
made of a collection of such filaments, streaming in random
directions. The motions in the large-linewidth central region, however, could
also partly be due to the energetics of the AGN that powers the central 
region.

There is no evidence of dust in the filaments. No extinction is seen
associated with the H$\alpha$ filaments, while dust associated with
the foreground high velocity system (at +3000km/s) is conspicuous
(Keel, 1983). The lack of significant amounts of dust in the filament,
presumably due to high sputtering rates, implies that the CO forms
mainly in the gaseous phase. Star formation activity is low. It is
only seen inside the filament in the Shields and Filipenko cluster. 
So the cold gas reservoir observed here is not forming stars very 
efficiently.

\section{Conclusions}

Our interferometer observations of the eastern filament region
reported in this paper show that we can now spatially resolve dense
cold gas along thin and elongated filaments coinciding with the
well-known H$\alpha$ emission around NGC1275. This cold gas is also
resolved into narrow linewidth-emitting clouds (30 km/s), in contrast
to the broad lines (200 to 400 km/s) detected in the majority of our
single dish detections (Edge et al, 2001, Salom\'e $\&$ Combes, 2003).
The interferometer retrieves all of the CO emission found in
single-dish observations, which implies that we have not missed any
more large-scale CO emission.  The east filament is resolved into
molecular gas concentrations that could be cooling separately.  The
molecular gas has probably formed in the spots where it is
observed. If the filament is on the rear side of the galaxy, then its
morphology, total mass, and kinematics is consistent with the scenario
proposed by Revaz et al.\ (2008) where the AGN feedback drags warm
gas to greater radii, where it undergoes runaway cooling.  This
process can create the east filament observed in CO, possibly composed
of GMCs partly falling toward the center of NGC~1275.

\begin{acknowledgements}
IRAM is supported by INSU/CNRS (France), MPG (Germany) and IGN
(Spain).  We thank the IRAM interferometer
operators for their expert help with the observing.
\end{acknowledgements}

\end{document}